\documentclass{aa}
\usepackage{txfonts}
\usepackage{graphicx}
\usepackage{longtable}
\usepackage{lscape}

\usepackage{natbib,twoopt}
\usepackage[breaklinks=true]{hyperref} 
\bibpunct{(}{)}{;}{a}{}{,} 
\makeatletter
\newcommandtwoopt{\citeads}[3][][]{\href{http://adsabs.harvard.edu/abs/#3}%
{\def\hyper@linkstart##1##2{}%
\let\hyper@linkend\@empty\citealp[#1][#2]{#3}}}
\newcommandtwoopt{\citepads}[3][][]{\href{http://adsabs.harvard.edu/abs/#3}%
{\def\hyper@linkstart##1##2{}%
\let\hyper@linkend\@empty\citep[#1][#2]{#3}}}
\newcommandtwoopt{\citetads}[3][][]{\href{http://adsabs.harvard.edu/abs/#3}%
{\def\hyper@linkstart##1##2{}%
\let\hyper@linkend\@empty\citet[#1][#2]{#3}}}
\newcommandtwoopt{\citeyearads}[3][][]%
{\href{http://adsabs.harvard.edu/abs/#3}
{\def\hyper@linkstart##1##2{}%
\let\hyper@linkend\@empty\citeyear[#1][#2]{#3}}}
\makeatother

\begin{document}

\title {Probing narrow-line Seyfert 1 galaxies in the southern hemisphere}
\author {
S. Chen \inst{1,2,3} \thanks{sina.chen@phd.unipd.it}
\and M. Berton \inst{1,4}
\and G. La Mura \inst{1}
\and E. Congiu \inst{1,4}
\and V. Cracco \inst{1}
\and \\ L. Foschini \inst{4}
\and J.H. Fan \inst{2}
\and S. Ciroi \inst{1,5}
\and P. Rafanelli \inst{1}
\and D. Bastieri \inst{3,6}
}

\institute {
Dipartimento di Fisica e Astronomia "G. Galilei", Universit\`a di Padova, Vicolo dell'Osservatorio 3, 35122, Padova, Italy;
\and Center for Astrophysics, Guangzhou University, 510006, Guangzhou, China;
\and Istituto Nazionale di Fisica Nucleare (INFN), Sezione di Padova, 35131, Padova, Italy;
\and INAF - Osservatorio Astronomico di Brera, Via E. Bianchi 46, 23807, Merate (LC), Italy;
\and INAF - Osservatorio Astronomico di Padova, Vicolo dell'osservatorio 5, 35122, Padova, Italy;
\and Dipartimento di Fisica e Astronomia "G. Galilei", Universit\`a di Padova, Via Marzolo 8, 35131, Padova, Italy.
}

\authorrunning{S. Chen et al.}
\titlerunning{}

\abstract {We present a new accurate catalog of narrow-line Seyfert 1 galaxies (NLS1s) in the southern hemisphere from the Six-degree Field Galaxy Survey (6dFGS) final data release, which is currently the most extensive spectroscopic survey available in the southern sky whose database has not yet been systematically explored. We classified 167 sources as NLS1s based on their optical spectral properties. We derived flux-calibrated spectra for the first time that the 6dFGS does not provide. By analyzing these spectra, we obtained strong correlations between the monochromatic luminosity at 5100 $\AA$ and the luminosities of H$\beta$ and [O III]$\lambda$5007 lines. The central black hole mass and the Eddington ratio have average values of $8.6 \times 10^{6} M_{\odot}$ and $0.96 L_{Edd}$ respectively, which are typical values for NLS1s. In the sample, 23 (13.8$\%$) NLS1s were detected at radio frequencies, and 12 (7.0$\%$) of them are radio-loud. Our results confirmed that radio-loud sources tend to have higher redshift, a more massive black hole, and higher radio and optical luminosities than radio-quiet sources.}


\maketitle

\section{Introduction}

Narrow-line Seyfert 1 galaxies (NLS1s) are a well-known subclass of active galactic nuclei (AGN). The classification is based on their optical spectral properties. Their H$\beta$ lines, originating in the broad-line region (BLR), have by definition a full width at half maximum FWHM (H$\beta$) $<$ 2000 km s$^{-1}$, and a flux ratio of [O III]$\lambda$5007 / H$\beta$ $<$ 3 \citep{Osterbrock1985, 1989ApJ...342..224G}. Most of their optical spectra show strong Fe II multiplets emission, which is a sign that the BLR and the accretion disk are visible \citep{2008MNRAS.385...53M, 2011nlsg.confE...2P}. These narrow permitted lines are commonly interpreted as a low rotational velocity around a relatively undermassive central black hole compared to broad-line Seyfert 1 galaxies (BLS1s), typically in the ranges of $M_{BH} \sim 10^{6-8} M_{\odot}$ for NLS1s and $M_{BH} \sim 10^{7-8} M_{\odot}$ for BLS1s \citep{2000MNRAS.314L..17M}. This leads to the conclusion that NLS1s have a high Eddington ratio, from 0.1 to 1 or even above \citep{1992ApJS...80..109B, 2004A&A...426..797C}. The low black hole mass and high accretion rate suggest that NLS1s might be a young and fast-growing phase of AGN \citep{2000MNRAS.314L..17M, 2000NewAR..44..455G}. In this scenario, NLS1s might be different with respect to BLS1s, as suggested by their different large-scale environments: NLS1s preferably reside in less dense environments than BLS1s \citep{2017A&A...606A...9J}. Another possible interpretation is that the narrowness of H$\beta$ lines in NLS1s might be due to the pole-on orientation of a disk-like BLR, which prevents us from seeing any Doppler broadening \citep{2008MNRAS.386L..15D}. In this case, the low black hole mass of NLS1s is not real, but only an inclination effect. However, our current knowledge is too poor to reach a conclusion.

Some NLS1s are detected at radio frequencies. The radio properties of AGN are usually described by the radio-loudness parameter, which is defined as the flux ratio between 5 GHz and optical B-band, $R_{L} = F_{5GHz} / F_{B-band}$. Sources with $R_{L} > 10$ are considered as radio-loud (RL), while the others belong to radio-quiet (RQ) population \citep{1989AJ.....98.1195K}. Among radio-detected NLS1s, the majority of them are RQ, while only a fraction of NLS1s are RL (7$\%$) or very RL (2.5$\%$) with a radio-loudness $R_{L} > 100$ \citep{2006AJ....132..531K}. The RL fraction of NLS1s is indeed half of the fraction of RL quasars (15$\%$) \citep{1989AJ.....98.1195K}. However, this fraction strongly depends on the redshift since it appears to be lower in the nearby universe \citep{Cracco2016}, and relies on the large-scale environment as it tends to be higher with increasing large-scale environment density \citep{2017A&A...606A...9J}. These RL NLS1s generally have a very compact radio morphology \citep{2010AJ....139.2612G, 2012ApJ...760...41D}. Instead, RQ NLS1s usually exhibit an extended emission \citep{2018arXiv180103519B}.

The discovery of $\gamma$-ray emission from NLS1s detected by the Fermi $\gamma$-ray Space Telescope confirmed that NLS1s are the third class of AGN emitting $\gamma$-rays from a relativistic beamed jet, in addition to blazars and radio galaxies \citep{Abdo2009b}. However only a handful of sources are detected at $\gamma$-rays. The physical properties of NLS1s are different from those of blazars and radio galaxies. The central black hole masses of NLS1s are on average one to two orders of magnitude lower than those of blazars and radio galaxies. This difference might be due to a projection effect in the inclination scenario. Moreover, NLS1s are generally hosted in spiral galaxies \citep{2003AJ....126.1690C}, while blazars and radio galaxies are both typically hosted in elliptical galaxies \citep{2007ApJ...658..815S}. If we assume that the young age scenario is correct, RL NLS1s might be the progenitors of flat spectrum radio quasars (FSRQs) \citep{Foschini2015, 2016A&A...591A..98B}.

To have a better understanding of the nature of NLS1s, observations with advanced observing facilities will be necessary. NLS1s are typically fainter with respect to blazars and quasars, and RL NLS1s are especially tricky since they are usually located at relatively high redshift. Some new facilities, such as the Extremely Large Telescope (ELT), the Atacama Large Millimeter/submillimeter Array (ALMA), and the Square Kilometre Array (SKA), could make a great contribution in providing break-through evidence to solve many long lasting problems concerning this particularly intriguing class of AGN. However, most of these large telescopes are concentrated in the southern hemisphere. Therefore the aim of this work is to create a new NLS1 sample that can be observed by large telescopes in the southern hemisphere, for the purpose of investigating the peculiarity of NLS1s with respect to other AGN. To do this, we exploited the large unexplored archive of the Six-degree Field Galaxy Survey (6dFGS)\footnote{http://www-wfau.roe.ac.uk/6dFGS/.}, and provided flux calibration to their spectra for the first time.

Throughout this work, we adopt a standard $\Lambda$CDM cosmology with a Hubble constant $H_{0}$ = 70 km s$^{-1}$ Mpc$^{-1}$, $\Omega_{\Lambda}$ = 0.73 and $\Omega_{M}$ = 0.27 \citep{2011ApJS..192...18K}. We assume the flux density and spectral index convention $ F_{\nu} \propto \nu^{-\alpha_{\nu}} $.

\section{Sample selection}

The spectral data of the NLS1 sample are derived from the 6dFGS, which is a combined redshift and peculiar velocity survey over the entire southern sky with $|b| > 10^{\circ}$. Optical spectra were obtained through separate V and R gratings and later spliced to produce combined spectra spanning 4000 $\AA$ - 7500 $\AA$. The instrumental resolution is R $\sim$ 1000 \citep{Jones2004}. The third and final data release (DR3) for the 6dFGS was published on 1 April 2009 \citep{Jones2009}.

We analyzed 110880 archive spectra from the 6dFGS DR3 catalog. They were first corrected for redshift using the $cz$ velocity parameter (km s$^{-1}$) provided by the catalog. The continuum was fitted in two wavelength intervals (4725 $\AA$ - 4775 $\AA$ and 5075 $\AA$ - 5125 $\AA$) that do not show strong emission line, and subtracted from the original spectra. The root mean square (RMS) was calculated in the regions mentioned above as well. We selected objects showing strong H$\beta$ and [O III] lines, based on the criterion that the peak of each emission line has to be larger than 3$\cdot$RMS. In this case, we found 38563 objects that having both emission lines.

\begin{figure}[ht]
\centering
\includegraphics[width=0.5\textwidth]{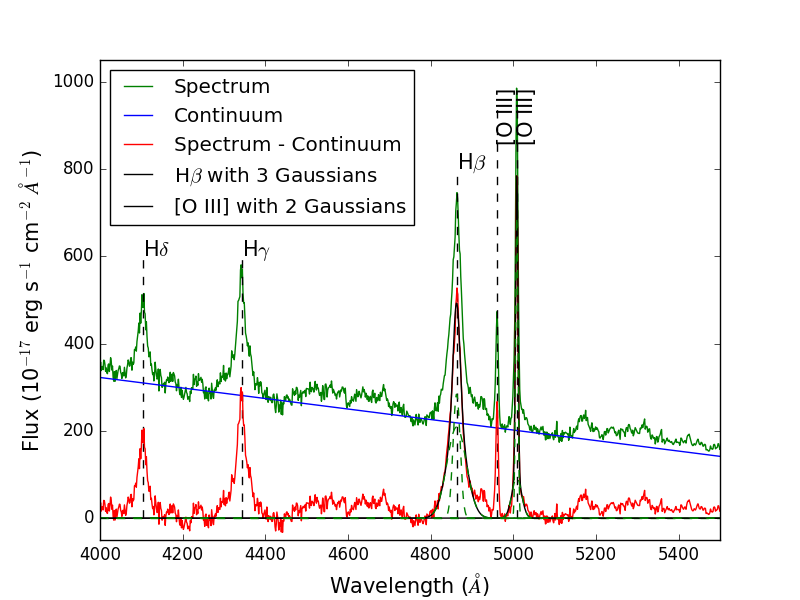}
\caption{\footnotesize Spectrum of 6dFGS gJ135439.5-421457. The green line is the observed spectrum, the blue line is the continuum, the red line is the spectrum after the continuum subtraction, and the black line is the result of fitting H$\beta$ with three Gaussians and [O III] with two Gaussians. Data are taken from the 6dFGS.}
\label{gaussian}
\end{figure}

We then calculated the FWHM of these emission lines. The H$\beta$ line was fitted with three Gaussians (one narrow component and two broad components), the [O III] line was fitted with two Gaussians (one narrow component and one broad component), and both of them were corrected for instrumental resolution (300 km s$^{-1}$ for the 6dFGS). The narrow component is the line core and the broad components are usually associated with the gas outflow. An example of continuum subtraction and emission lines Gaussian fitting is shown in Fig. ~\ref{gaussian}.

After reproducing the line profiles, we applied two additional conditions on the FWHM of lines. The first one was 600 km s$^{-1}$ $<$ FWHM(H$\beta$) $<$ 2200 km s$^{-1}$. The lower limit is based on the mean FWHM of emission lines from the narrow-line region (NLR) measured by \citet{2012MNRAS.427.1266V}, and the upper limit was defined by the classic 2000 km s$^{-1}$ criterion with a 10 $\%$ tolerance \citep{Cracco2016}. The second condition was 200 km s$^{-1}$ $<$ FWHM([O III]) $<$ 2000 km s$^{-1}$. The lower limit avoids contamination by cosmic rays, while the upper limit avoids loss of objects with strong Fe II multiplets and relatively weak [O III] line \citep{Osterbrock1985, 2013MNRAS.433..622M}. The fluxes of the H$\beta$ and [O III] lines were measured by integrating the fitted profiles. We only focused on those objects with a flux ratio of [O III] / H$\beta$ $<$ 3, which is a main feature of NLS1s \citep{Osterbrock1985}.

\begin{figure}[ht]
\centering
\includegraphics[width=0.5\textwidth]{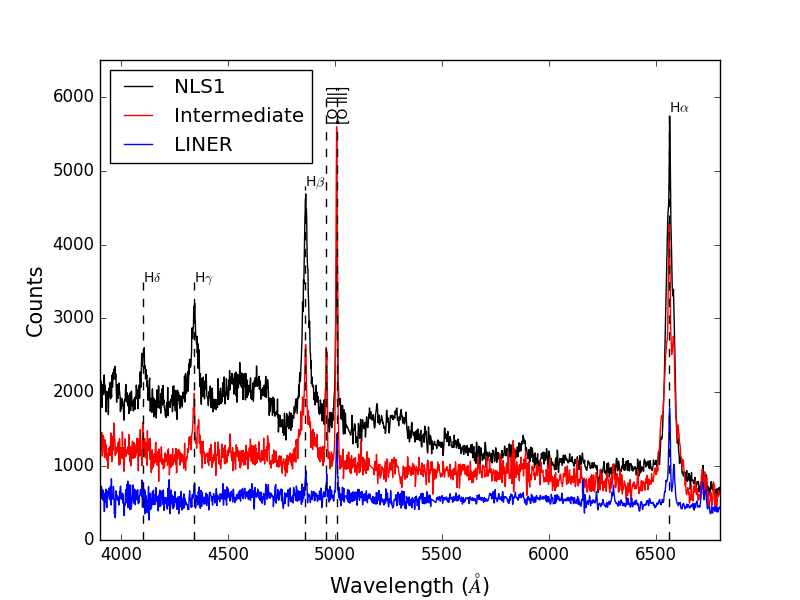}
\caption{\footnotesize Three spectral examples showing the differences between a NLS1 6dFGS gJ123124.9-165350 (black line), an intermediate Seyfert galaxy 6dFGS gJ072957.1-654333 (red line), and a LINER 6dFGS gJ160558.2-263806 (blue line). Data are taken from the 6dFGS.}
\label{compare}
\end{figure}

Based on the criteria mentioned above, we selected 2126 spectra and further analyzed them one by one to discern NLS1s from intermediate Seyfert galaxies and low-ionization nuclear emission-line regions (LINERs). Intermediate Seyfert galaxies and LINERs also have narrow emission lines and a flux ratio of [O III] / H$\beta$ $<$ 3 \citep{Osterbrock1977, 1987ApJS...63..295V}. However, there are two features in the spectra of NLS1s that distinguish them from the others. On one hand, Fe II multiplets are usually present in the spectra, because the BLR, where Fe II multiplets are originated, is directly visible in type 1 AGN. On the other hand, the line profile of H$\beta$ is broader than [O III] and can be fitted with a Lorentzian profile caused by turbulence in the line emitting region \citep{2013A&A...549A.100K}. The differences between NLS1s, intermediate Seyfert galaxies, and LINERs can be seen in Fig. ~\ref{compare}.

\begin{figure}[ht]
\centering
\includegraphics[width=0.5\textwidth]{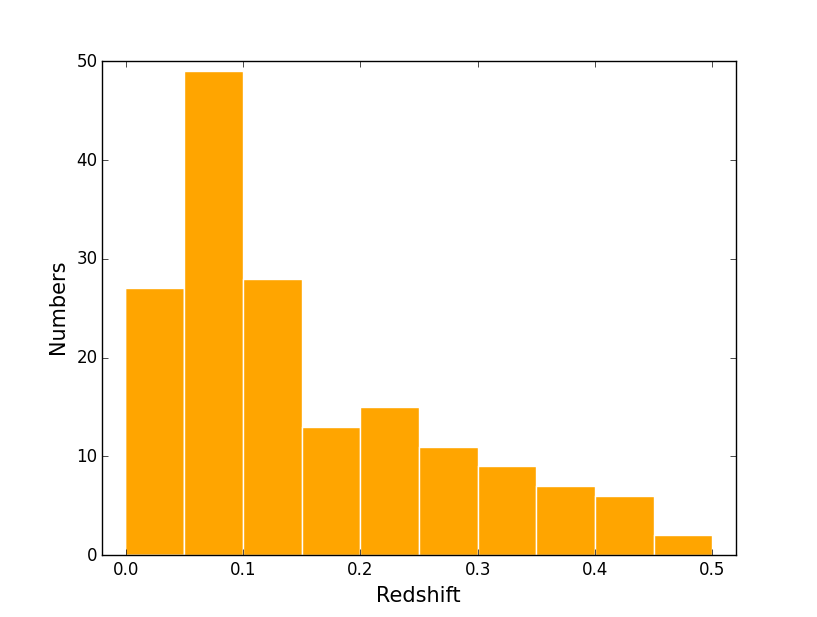}
\caption{\footnotesize Redshift distribution of 167 NLS1s using a 0.05 bin width.}
\label{redshift}
\end{figure}

In this way, we obtained a new sample of 167 NLS1s in the southern hemisphere. The object's name, coordinates, redshift, and luminosity distance are reported in Table ~\ref{All1}; FWHM and flux of H$\beta$ and [O III] lines are listed in Table ~\ref{All2}. The range of redshift is from $z$ = 0.01 ($D_{L}$ = 44.49 Mpc) to $z$ = 0.50 ($D_{L}$ = 2825.63 Mpc) with an average value of $z$ = 0.15 ($D_{L}$ = 767.40 Mpc). The redshift distribution of the sample is presented in Fig. ~\ref{redshift}.

\setcounter{table}{2}

\section{Flux calibration}

Spectra from the 6dFGS are not flux-calibrated, therefore a method to calibrate the spectra in the sample is needed. We looked for NLS1s that are both in the 6dFGS sample and observed by the Sloan Digital Sky Survey (SDSS), based on a coordinates match between the 6dFGS and the SDSS spectroscopic database, within a tolerance radius of 0.1 arcmin. The images of the sources in both surveys were compared to make sure that the search resulted in the same sources that we were looking for. As a result, seven NLS1s observed by both the 6dFGS and the SDSS were selected for the flux calibration.

We also checked that these objects are not RL to reduce the effects of optical variability. This is because RL sources are found to have higher variability than RQ sources due to the presence of non-thermal emission coming from the relativistic jet along with thermal emission related to the accretion disk around the supermassive black hole in RL objects, compared to the contribution of only thermal emission from the accretion disk in RQ objects \citep{2017ApJ...842...96R}.

\begin{figure}[ht]
\centering
\includegraphics[width=0.5\textwidth]{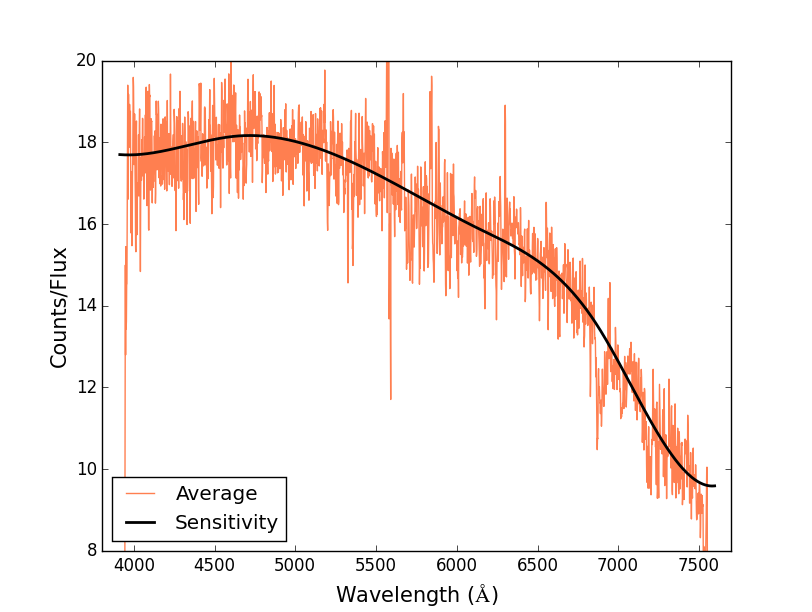}
\caption{\footnotesize Average $counts / flux$ ratio (red line) and sensitivity (black line) of the 6dFGS NLS1 sample.}
\label{sens}
\end{figure}

We calculated the ratio between counts and flux for each object, dividing the 6dFGS spectra by the SDSS spectra, then combined the results and averaged them using the IRAF V2.16. The sensitivity of the 6dFGS NLS1 sample was obtained by fitting the average ratio with a fifth order polynomial curve. The average $counts / flux$ ratio and the sensitivity are shown in Fig. ~\ref{sens}. The airmass was considered in a crude correction for all data using the standard stars Feige 110 and EG274 in the 6dFGS observations \citep{Jones2004}. Hence, we calibrated the NLS1 spectra in the sample using the sensitivity obtained above. The spectra of these 167 NLS1s were then corrected for galactic extinction using the $A_V$ extinction coefficients (Landolt V) \citep{2011ApJ...737..103S} extracted from the NASA/IPAC \footnote{The National Aeronautics and Space Administration / The Infrared Processing and Analysis Center.} Extragalactic Database (NED) \footnote{https://ned.ipac.caltech.edu/.}, and shifted to the rest frame. In this way we obtained flux-calibrated spectra for the NLS1 sample.

An issue with the method we used to calibrate the spectra is that the sensitivity was calculated on a limited number of objects, because of the small common sky coverage of the 6dFGS and the SDSS. In order to examine whether the flux calibration is acceptable, we compared the relation between the continuum flux and the optical magnitude of our 6dFGS re-calibrated spectra with that of NLS1s flux-calibrated spectra selected by \citet{Cracco2016}, who investigated a sample of 296 NLS1s from the SDSS Data Release (DR) 7.

We performed averaged flux measurements running on the wavelength range from 5050 $\AA$ to 5150 $\AA$, to obtain the mean flux at 5100 $\AA$ for both the 6dFGS and SDSS samples. The magnitude on B-band of the 6dFGS sample is provided by the catalog. The SDSS, instead, gives the magnitude of photometric measurements in the $ugriz$ system. Hence, we converted the magnitude on $g$ and $r$ bands for the objects in the SDSS sample into magnitude on B-band using the equation \citep{2005AJ....130..873J}
\begin{equation}
B = 1.28 g - 0.28 r + 0.09.
\end{equation}
The magnitudes were also corrected for galactic extinction in both samples.

\begin{figure}[ht]
\centering
\includegraphics[width=0.5\textwidth]{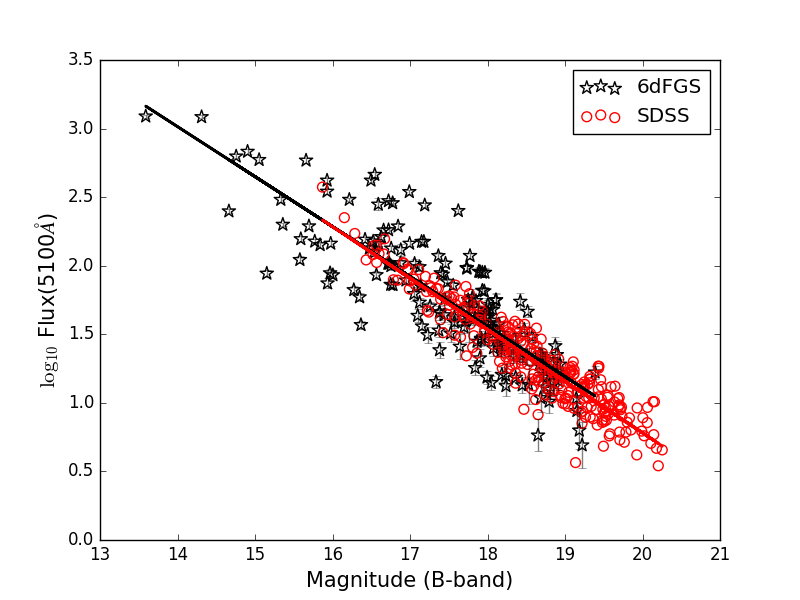}
\caption{\footnotesize Relation of the mean flux at 5100 $\AA$ and the magnitude on B-band. The black stars and line are data from the 6dFGS sample, the red circles and line are data from the SDSS sample in Cracco et al. (2016).}
\label{flux+mag}
\end{figure}

The relation between the mean flux at 5100 $\AA$ and the magnitude on B-band for the 6dFGS and SDSS samples is shown in Fig. ~\ref{flux+mag}. The best fit for the 6dFGS sample is
\begin{equation}
\log{F(5100 \AA)} = -(0.365 \pm 0.007) B + (8.117 \pm 0.147),
\end{equation}
which is comparable with the one for the SDSS sample expressed by
\begin{equation}
\log{F(5100 \AA)} = -(0.376 \pm 0.003) B + (8.298 \pm 0.067).
\end{equation}
Based on the consistent properties of our re-calibrated 6dFGS spectra with respect to the full flux-calibrated SDSS spectra, we infer that the derived sensitivity provides a realistic solution, though introducing some degrees of uncertainty. The magnitude on B-band and the mean flux density at 5100 $\AA$ of the 6dFGS NLS1 sample are listed in Table ~\ref{All1}.

We carried out an additional test for our flux calibration. Sources with a declination higher than $-15^{\circ}$ and apparent magnitude brighter than 18 and 19 can be observed by the Asiago Astrophysical Observatory (Italy) 1.22 m and 1.82 m telescopes respectively. We observed 15 objects' optical spectra in the 6dFGS sample using these telescopes in October and November 2017 in the best seeing conditions of 1" (details are reported in Table ~\ref{observe}). The observations were split into exposure times of 1200 s or 1800 s for each target to avoid strong contamination by cosmic rays and light pollution. The wavelength calibration was done by FeAr or NeHgAr lamps. Standard stars HR7596, HR9087, or HD2857, whose altitude close to the target, was used for the flux calibration. After the sky subtraction, we combined the spectra for each object, and corrected for galactic absorption and redshift.

\begin{table}
\caption{Observational details of the Asiago optical spectra.}
\label{observe}
\centering
\resizebox{0.5\textwidth}{!}{
\begin{tabular}{cccc}
\hline
\hline
Name & Observed date & Telescope & Exposure time \\
- & (yyyy-mm-dd) & - & (s) \\
\hline
6dFGS gJ000040.3-054101 & 2017-10-15 & 1.82 m & 7200 \\
6dFGS gJ002249.2-103956 & 2017-10-13 & 1.82 m & 3600 \\
6dFGS gJ013809.5-010920 & 2017-10-14 & 1.82 m & 5400 \\
6dFGS gJ015930.7-112859 & 2017-10-16 & 1.82 m & 5400 \\
6dFGS gJ020349.0-124717 & 2017-10-15 & 1.22 m & 7200 \\
6dFGS gJ021218.3-073720 & 2017-10-14 & 1.22 m & 7200 \\
6dFGS gJ021355.1-055121 & 2017-10-16 & 1.82 m & 5400 \\
6dFGS gJ034713.9-132547 & 2017-10-14 & 1.82 m & 3600 \\
6dFGS gJ042021.7-053054 & 2017-10-15 & 1.82 m & 5400 \\
6dFGS gJ043622.3-102234 & 2017-11-14 & 1.22 m & 4800 \\
6dFGS gJ044720.7-050814 & 2017-10-16 & 1.22 m & 3600 \\
6dFGS gJ045557.5-145641 & 2017-10-16 & 1.82 m & 3600 \\
6dFGS gJ193733.0-061305 & 2017-10-14 & 1.22 m & 4800 \\
6dFGS gJ211524.9-141706 & 2017-10-15 & 1.22 m & 4800 \\
6dFGS gJ213632.0-011626 & 2017-10-16 & 1.82 m & 5400 \\
\hline
\end{tabular}
}
\end{table}

\begin{figure}[ht]
\centering
\includegraphics[width=0.5\textwidth]{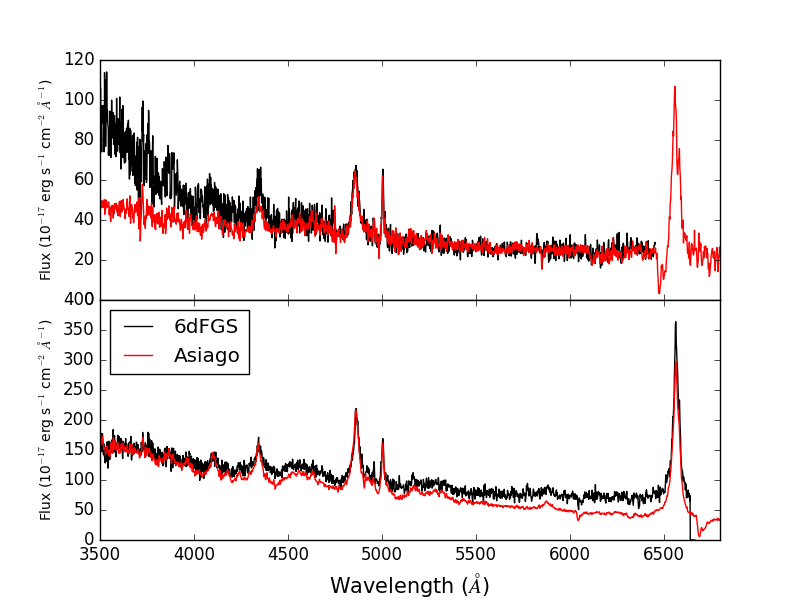}
\caption{\footnotesize Top panel: Flux-calibrated spectra of 6dFGS gJ021218.3-073720 obtained from the 6dFGS (black line) and the Asiago 1.22 m Telescope (red line). Bottom panel: Flux-calibrated spectra of 6dFGS gJ045557.5-145641 obtained from the 6dFGS (black line) and the Asiago 1.82 m Telescope (red line).}
\label{Asiago}
\end{figure}

We compared the flux-calibrated spectra obtained from the 6dFGS and the Asiago telescopes. Two examples are shown in Fig. ~\ref{Asiago}. In spite of the fact that the sources used to create the sensitivity are not RL, we do expect some variations and the most critical effect on the thermal blue bump. On the other hand, differences on the red part can be well accounted for by the different aperture and seeing conditions. Thus even though the uncertainties on the blue and red extremes of the spectral range are not negligible, the fluxes around the H$\beta$ line, [O III] line, and 5100 $\AA$ continuum regions are consistent. Hence we suppose that for our purpose the flux calibration result is reliable, and these regions can be used to calculate the black hole mass and the Eddington ratio.

\section{Luminosity correlation}

Working with spectra produced by a multi-object spectrograph, and fed by a system of fixed aperture entrance fibers, we have to account for the amount of light that enters the instrument. On one hand, we think about the nuclear active regions of external galaxies as unresolved point-like sources, the aperture of the 6dFGS instrument fibers is large enough to avoid the loss of significant amounts of light. On the other hand, the use of a fixed aperture implies that a varying amount of light from the AGN host galaxy, which can be an extended source with resolved structure, is also collected by the instrument. For strong type 1 AGN, which dominate over the host, this problem is not particularly relevant, but it can represent a significant issue in the case of nearby resolved Seyfert galaxies \citep{Varisco2018}. In general, this problem does not affect the spectral components that are purely originated by the AGN, such as the broad emission lines, while we have to take it into account when measuring the intensity of continuum, and to some extent, the narrow emission lines.

In fact, estimating the host galaxy contribution in such spectra is not straightforward, because the relative amount of host galaxy stellar light with respect to the AGN signal, depends both on the AGN luminosity and on the fraction of the host surface that is subtended by the aperture that increases as a square of distance. To evaluate the contribution of stellar light from host galaxies in our sample, we applied a technique based on the principal component analysis (PCA), which was originally described in \citet{Connolly1995}, and subsequently applied to a sample of type 1 AGN selected from the SDSS DR3 by \citet{LaMura2007}. Following the strategy described in \citet{LaMura2007}, we assume that the observed spectra can be regarded as the linear combination of a set of spectral contributions, arising separately from the AGN and its host. If we call $S(\lambda)$ the amount of flux observed at a wavelength $\lambda$, this leads us to write
\begin{equation}
S(\lambda) = \sum_{i = 1}^n a_i A_i(\lambda) + \sum_{j = 1}^m h_j H_j(\lambda), \label{eqPCA}
\end{equation}
where $A_i(\lambda)$ and $H_j(\lambda)$ represent the spectral components produced by the AGN and its host respectively, $a_i$ and $h_j$ are their weight coefficients that do not depend on wavelength, and $n$ and $m$ are the number of linear components used to model the AGN and its host. Using a proper set of base spectra, such as the list of eigenspectra provided by the low redshift collections in \citet{Yip2004a, Yip2004b}, Equation ~(\ref{eqPCA}) expands to a system of algebraic equations that can be solved in terms of the weight coefficients. Therefore we can estimate the contribution of AGN and host galaxy components in the spectra, after they have been corrected for galactic extinction and brought to their rest frame.

\begin{figure}[ht]
\centering
\includegraphics[width=0.5\textwidth]{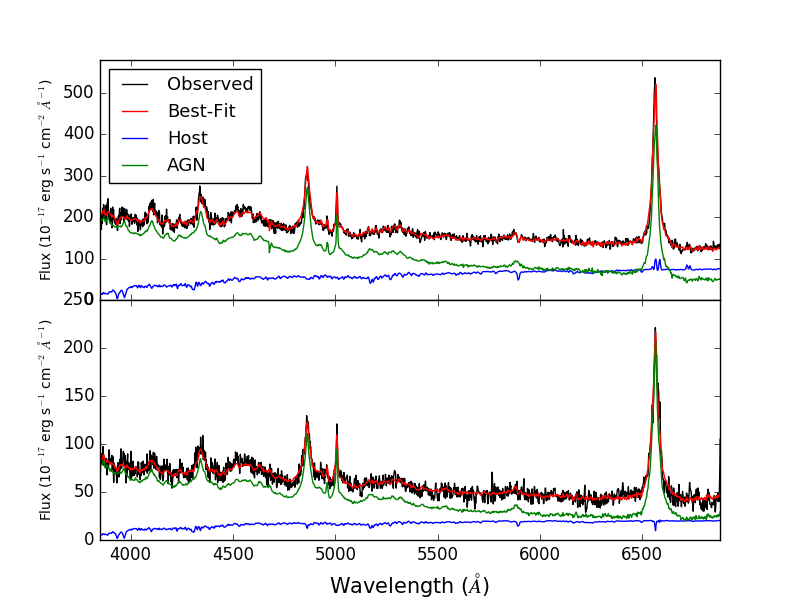}
\caption{\footnotesize Decompositions of AGN and host galaxy in spectra 6dFGS gJ065017.5-380514 (top panel) and 6dFGS gJ084510.2-073205 (bottom panel). The black line shows the observed spectrum, the red line shows the best-fitted result, the green line shows the AGN component, and the blue line shows the stellar contribution.}
\label{agn+host}
\end{figure}

As the simple solution of Equation ~(\ref{eqPCA}) may not always lead to a physically meaningful conclusion, because it introduces negative spectra or absorption features corresponding to forbidden lines, we developed an iterative fitting procedure written in Python. The algorithm looks for the solution of Equation ~(\ref{eqPCA}), then it tests for the physical meaning of the separated AGN and host spectra. If either one of the components exhibits any inconsistency, its weight is reduced and the fit is run again leaving only the consistent parts free to vary. The algorithm then proceeds with an iteration fitting AGN and host alternatively, which are now forced to show only non-negative contributions, until the solution converges to a minimum chi-square $\chi^2$ value or one of the components becomes undetected. After testing various configurations, we found that the best fit models leaving minimum residuals could be obtained from the combination of the first $n$ AGN and the first $m$ host galaxy components. Two examples of AGN and host galaxy decompositions are shown in Fig. ~\ref{agn+host}.

The procedure is generally able to recover the host galaxy contribution, producing models with a median reduced residual of $\chi^{2} = 1.45$ estimated over the sample. In a dozen cases, the host contribution turned out to be too faint to be distinguished. In all the other cases, we subtracted the host galaxy spectral models from our flux-calibrated spectra, in order to obtain an estimate of the light originally emitted by the AGN alone. After the host galaxy correction, the monochromatic luminosity at 5100 $\AA$ can be calculated by
\begin{equation}
L(\lambda) = 4 \pi D_{L}^{2} \cdot \lambda F(\lambda),
\end{equation}
where $D_{L}$ is the luminosity distance estimated by the cosmological redshift, $F(\lambda)$ is the averaged flux density on the wavelength range from 5050 $\AA$ to 5150 $\AA$, and we assume that the radiation from the sources is isotropic. The luminosity at 5100 $\AA$ may be overestimated in RL NLS1s due to the presence of relativistic jets can contaminate the continuum in RL sources. However, this impact is negligible, since only a small fraction of NLS1s are RL as mentioned above.

\begin{figure}[ht]
\centering
\includegraphics[width=0.5\textwidth]{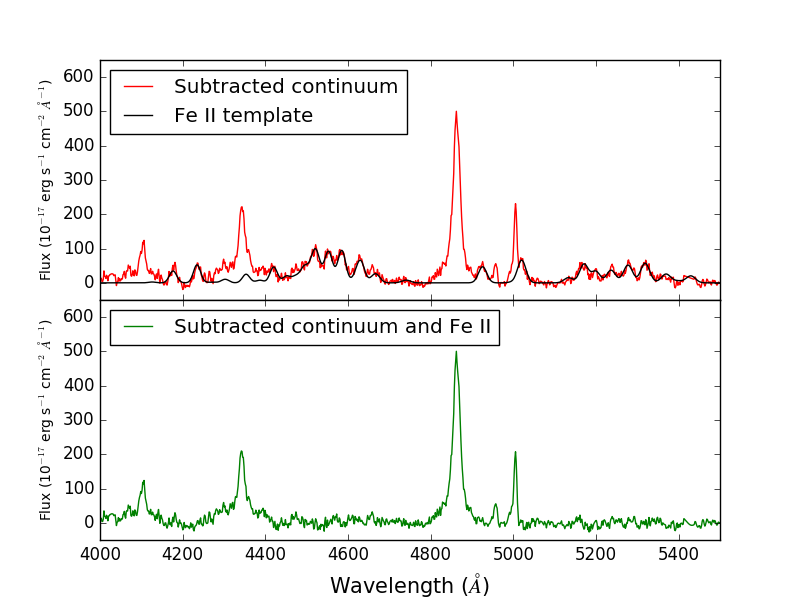}
\caption{\footnotesize The Fe II multiplets fitting in spectrum 6dFGS gJ003915.9-511702. Top panel: The continuum-subtracted spectrum (red line) and the Fe II template (black line). Bottom panel: The spectrum after subtraction of both continuum and Fe II multiplets (green line). Data are taken from the 6dFGS.}
\label{FeII}
\end{figure}

From the host galaxy subtractions, we noted that the stellar light has an important influence on the continuum, however, it only makes a minor contribution to the emission lines, since we did not see any strong emission or absorption line in the host galaxy components. Therefore, the luminosities of H$\beta$ and [O III] lines can be estimated based on the flux-calibrated spectra. We first subtracted the continuum from the spectra, then fitted the Fe II multiplets and subtracted them from the continuum-subtracted spectra. The Fe II templates were estimated using the procedure described in \citet{2010ApJS..189...15K, 2012ApJS..202...10S}. It reproduces 65 Fe II emission lines within the 4000 $\AA$ to 5500 $\AA$ range, including five line groups (F, S, G, P, I Zw 1), and fits each line with a single Gaussian. The free input parameters are temperature, Doppler width, shift, intensity of each group of lines, and iterations \footnote{http://servo.aob.rs/FeII$\_$AGN/.}. An example of Fe II multiplets fitting is shown in Fig. ~\ref{FeII}.

We used the spectra with both the continuum and the Fe II multiplets subtracted to estimate the fluxes and luminosities of H$\beta$ and [O III] lines. The signal-to-noise (S/N) ratio was computed from 5050 $\AA$ to 5150 $\AA$ wavelength range. If S/N $>$ 10, the H$\beta$ line was fitted with three Gaussians (one narrow component and two broad components). Otherwise it was fitted with two Gaussians (one narrow component and one broad component), in order to avoid overfitting the noise in the case of low quality spectra. The [O III] line was fitted with two Gaussians. The luminosities of H$\beta$ and [O III] lines is calculated by
\begin{equation}
L(line) = 4 \pi D_{L}^{2} \cdot F(line),
\end{equation}
where $F(line)$ is the flux derived by integrating the H$\beta$ and [O III] line profiles.

\begin{figure}[ht]
\centering
\includegraphics[width=0.5\textwidth]{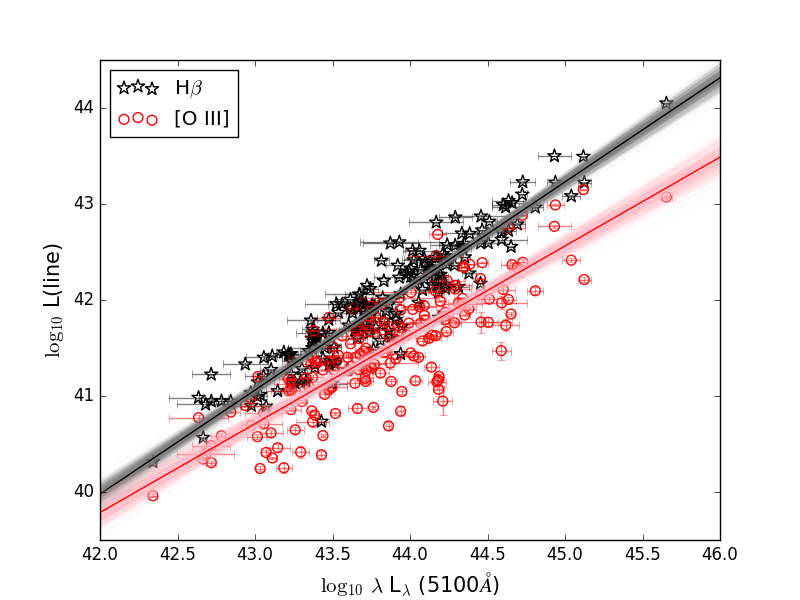}
\caption{\footnotesize Relation of L(H$\beta$) - $\lambda$L$_{\lambda}$(5100$\AA$) is shown in black stars and lines. Relation of L([O III]) - $\lambda$L$_{\lambda}$(5100$\AA$) is shown in red circles and lines.}
\label{luminosity}
\end{figure}

The luminosity at 5100 $\AA$, and the luminosities of H$\beta$ and [O III] lines for the 6dFGS NLS1 sample are reported in Table ~\ref{All3}. We obtained strong correlations between the monochromatic luminosity at 5100 $\AA$ and the luminosities of both emission lines as shown in Fig. ~\ref{luminosity}. The linear regressions with a Bayesian method yield \citep{Kelly2007}
\begin{equation}
\log{L(H\beta)} = (1.084 \pm 0.031) \log{\lambda L_{\lambda} (5100 \AA)} - (5.560 \pm 1.375),
\end{equation}
and
\begin{equation}
\log{L([O III])} = (0.926 \pm 0.047) \log{\lambda L_{\lambda} (5100 \AA)} + (0.909 \pm 2.058).
\end{equation}

A strong correlation between the luminosity of Balmer lines and the optical continuum was found by \citet{2005ApJ...630..122G, 2010ApJ...723..409G}, and supports the idea that the Balmer emission lines arise from recombination after photoionization by non-thermal continuum \citep{1980ApJ...241..894Y}. \citet{2006ApJS..166..128Z} also found a correlation between the luminosity of [O III] line and the optical continuum, although not as tight as the L(H$\beta$) - $\lambda$L$_{\lambda}$(5100$\AA$) correlation. This relation is useful for estimating the intrinsic luminosity of AGN, since the unobscured NLR, where the [O III] line comes from, can be reliably used to trace the central region properties \citep{2003AJ....126.2125Z}.

The errors of flux and luminosity of continuum were obtained by calculating their standard deviation $\sigma$ from 5050 $\AA$ to 5150 $\AA$. The errors of FWHM, flux, and luminosity of emission lines were evaluated by the Monte Carlo method. This approach involves varying the line profiles with a random noise proportional to the RMS measured in the continuum range, then fitting the emission lines with Gaussians as previously mentioned, measuring FWHM, flux, and luminosity, and repeating the same process 100 times to estimate their standard deviation. In this way, we obtained $1 \sigma$ errors for these parameters.

\section{Black hole mass and Eddington ratio}

We estimated the central black hole mass following the method in \citet{Foschini2015, Berton2015}. For each NLS1 galaxy in the 6dFGS sample, the virial mass of the central black hole is defined as
\begin{equation}
M_{BH} = f \left( \frac{R_{BLR} \sigma^{2}_{line}}{G} \right),
\end{equation}
where $R_{BLR}$ is the size of the BLR measured by reverberation or estimated from scaling relations, $\sigma_{line}$ is the line dispersion second-order moment, $G$ is the gravitational constant, and $f$ is a dimensionless scale factor that depends on the structure, kinematics, and orientation of the BLR \citep{2004ApJ...613..682P}. We assumed $f = 4.31 \pm 1.05$ following \citet{Grier2013}, which is a value not dependent on the inclination and geometry of the BLR.

The size of the BLR was calculated using the relation between the luminosity at 5100 $\AA$ and the radius of the BLR from \citet{Bentz2013},
\begin{equation}
\log{\left(\frac{R_{BLR}}{1 \, \rm{light \, day}}\right)} = 1.527^{+0.031}_{-0.031} + 0.533^{+0.035}_{-0.033} \log{\left(\frac{\lambda L_{\lambda} (5100 \AA)}{10^{44} \, \rm{erg \, s^{-1}}}\right)}.
\end{equation}
The line dispersion $\sigma_{H\beta}$, which is less affected by the inclination and line profile than a FWHM measurement \citep{2004ApJ...613..682P, 2006A&A...456...75C}, was determined from the H$\beta$ broad components. If the H$\beta$ line has two broad components, the $\sigma_{H\beta}$ is evaluated by
\begin{equation}
\sigma_{H\beta}^{2} = \frac{\int \lambda^{2} F(\lambda) d\lambda}{\int F(\lambda) d\lambda} - \left( \frac{\int \lambda F(\lambda) d\lambda}{\int F(\lambda) d\lambda} \right)^{2},
\end{equation}
where $F(\lambda)$ is the H$\beta$ line profile produced by two broad components. If it has only one broad component, then $\sigma_{H\beta}$ $\simeq$ FWHM$_{H\beta_{broad}}$ / 2.355.

The Eddington ratio, defined as $\epsilon = L_{bol} / L_{Edd}$, could be measured after obtaining the central black hole mass, where the Eddington luminosity is
\begin{equation}
L_{Edd} = 1.3 \times 10^{38} \left(\frac{M_{BH}}{M_{\odot}}\right) \, \rm{erg \, s^{-1}},
\end{equation}
and the bolometric luminosity is estimated assuming $L_{bol} = 9 \lambda L_{\lambda} (5100\AA)$ \citep{2000ApJ...533..631K}.

\begin{figure}[ht]
\centering
\includegraphics[width=0.5\textwidth]{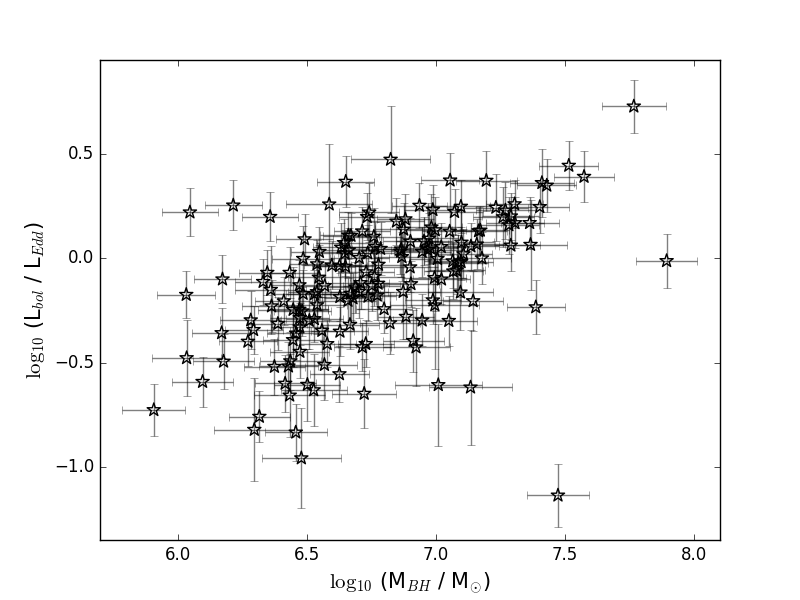}
\caption{\footnotesize Distributions of central black hole mass and Eddington ratio in the 6dFGS NLS1 sample.}
\label{Mass+Eddington}
\end{figure}

The distributions of central black hole mass and Eddington ratio for 167 NLS1s in the 6dFGS sample are plotted in Fig. ~\ref{Mass+Eddington}, and their estimated values are presented in Table ~\ref{All3}. The masses of the central black hole $M_{BH}$ range from $8.1 \times 10^{5} M_{\odot}$ to $7.8 \times 10^{7} M_{\odot}$ with a median value of $8.6 \times 10^{6} M_{\odot}$. The Eddington ratios $L_{bol} / L_{Edd}$ span from 0.07 to 5.35 with a median value of 0.96. This result confirms that NLS1s have lower black hole mass and higher Eddington ratio than BLS1s. The errors of central black hole mass and Eddington ratio were evaluated by the Python uncertainties package following the error propagation theory.

\section{Radio sources}

With the aim of looking for radio sources in the 6dFGS sample, we subsequently cross-matched these 167 NLS1s with several important radio surveys covering the southern hemisphere, including the NRAO VLA \footnote{The National Radio Astronomy Observatory Very Large Array.} Sky Survey (NVSS), the Sydney University Molonglo Sky Survey (SUMSS), and the Australia Telescope 20 GHz Survey (AT20G), within a search radius of 5 arcsec. The NVSS covers the entire sky north of $-40^{\circ}$ declination at 1.4 GHz with the flux density limit of 2.5 mJy \citep{1998AJ....115.1693C}, the SUMSS covers the sky south of $-30^{\circ}$ declination at 843 MHz with the flux density limit of 8 mJy \citep{2003MNRAS.342.1117M}, and the AT20G covers the whole sky south of $0^{\circ}$ declination at 5 GHz, 8.4 GHz, and 22 GHz respectively, with the flux density limit of 40 mJy \citep{2010MNRAS.402.2403M}. 

We found 18 sources only detected by the NVSS, three sources only detected by the SUMSS, one source detected by both the NVSS and the SUMSS, and one source detected by both the SUMSS and the AT20G. In total 23 (13.8$\%$) sources have associated radio counterparts. The flux densities detected by the radio surveys for these 23 sources are listed in Table ~\ref{radio} respectively.

The radio loudness $R_{L} = F_{radio} / F_{optical}$ was calculated using the radio flux at 5 GHz and the optical flux at B-band 4400 $\AA$ for each object. In most cases the radio flux was obtained at 1.4 GHz and 843 MHz instead of 5 GHz. We derived the radio flux at 5 GHz under the hypothesis that the radio spectrum can be described with a power-law model $F_{\nu} \propto \nu^{-\alpha}$, and we assumed a conservative spectral index of $\alpha = 0.5$ \citep{2008ApJ...685..801Y}. In this way we found 12 (7.0$\%$ of the whole sample) RL (including one very RL) NLS1s and 11 RQ NLS1s. If we suppose that the radio spectrum has a flat spectral index of $\alpha = 0$, 17 objects were found to be RL and six objects were found to be RQ. Instead, assuming a steep radio spectral index of $\alpha = 1$, we found that seven objects are RL and 16 objects are RQ. In this paper, we classified our sources according to the results of $\alpha = 0.5$.

\citet{2002AJ....124.3042W} analyzed 150 NLS1s in the SDSS Early DR and found that only a dozen (8$\%$) were detected at radio frequencies and only two (1.3$\%$) were RL. Research by \citet{2006ApJS..166..128Z} based on the SDSS DR3 resulted in a sample of 2011 NLS1s and of those 142 (7.1$\%$) objects had radio counterparts. \citet{Cracco2016} investigated a sample of 296 NLS1s from the SDSS DR7, 70 (23.6$\%$) sources were detected at radio frequencies and 11 (3.7$\%$ of the total sample and 15.7$\%$ of those radio detected) of them were classified as RL. A recent study by \citet{2017ApJS..229...39R} proposed a new catalog from the SDSS DR12 that contains 11101 NLS1s, among them 555 (5$\%$) objects were detected on radio band and 378 (3.4$\%$) were RL.

The radio fraction is different in all these samples because the radio loudness strongly depends on the spatial resolution of the optical and radio observations. The flux is different if the observations considered the whole galaxy (including the contribution of both AGN and host) or only the nuclear region. Then the evaluated $R_{L}$ is different for the same galaxy. This situation happens both on radio and optical observations. \citet{2001ApJ...555..650H} found that most Seyfert 1 nuclei were classified as RL AGN if only their nuclear luminosity was considered, which was in doubt later.

The classification of RL and RQ according to $R_{L}$ = 10 is not restricted and a little arbitrary. In light of this, a recent work by \citet{Padovani2017} argued that AGN should be classified based on physical differences rather than just an observational phenomenon. They defined two new classes of AGN: jetted AGN, which are characterized by strong relativistic jets, and non-jetted AGN, which also have jet structures but with weak power compared to those of jetted sources. The spectral energy distribution (SED) of non-jetted AGN has a cutoff at much lower energy than jetted AGN. Despite this, the radio-loudness parameter has been useful to separate RL sources from RQ ones until now.

\begin{figure}[ht]
\centering
\includegraphics[width=0.5\textwidth]{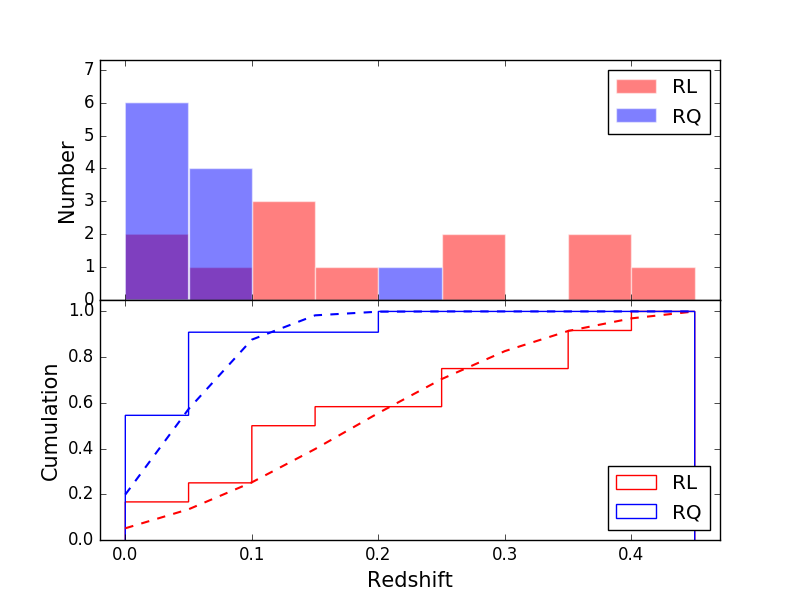}
\caption{\footnotesize Redshift distribution (top panel) and cumulative distribution (bottom panel) of the RL (red) and RQ (blue) sources in our radio detected NLS1 sample using a 0.05 bin width.}
\label{radio+z}
\end{figure}

\begin{figure}[ht]
\centering
\includegraphics[width=0.5\textwidth]{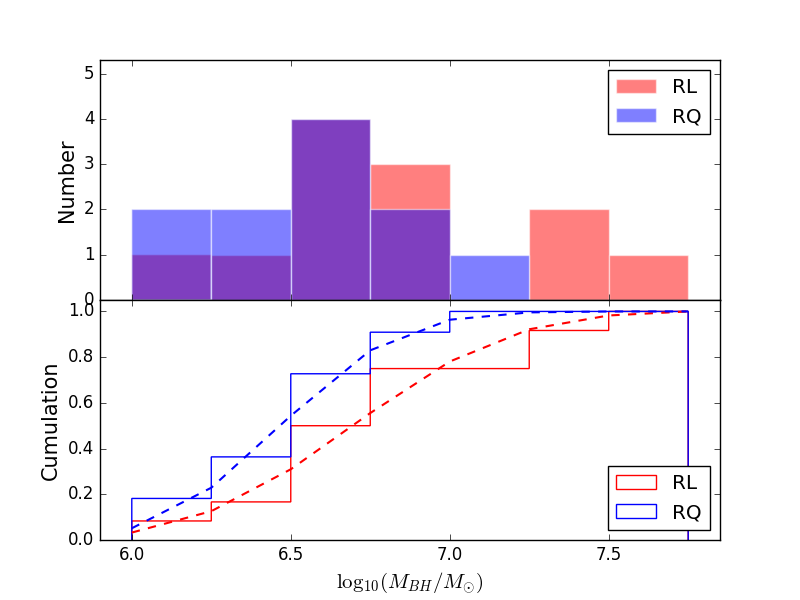}
\caption{\footnotesize Black hole mass distribution (top panel) and cumulative distribution (bottom panel) of the RL (red) and RQ (blue) sources as in previous figure using a 0.25 bin width.}
\label{radio+BHmass}
\end{figure}

\begin{figure}[ht]
\centering
\includegraphics[width=0.5\textwidth]{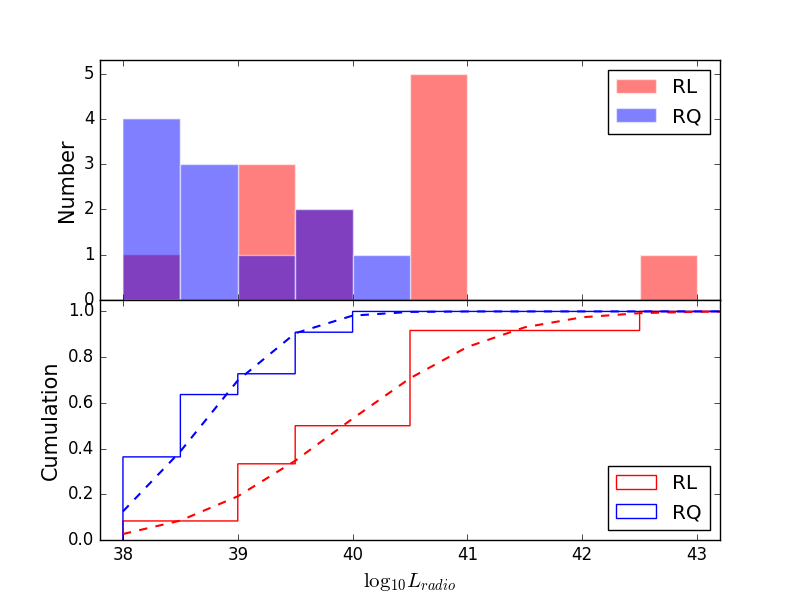}
\caption{\footnotesize Radio luminosity at 5 GHz distribution (top panel) and cumulative distribution (bottom panel) of the RL (red) and RQ (blue) sources as in previous figure using a 0.5 bin width.}
\label{radio+LR}
\end{figure}

\begin{figure}[ht]
\centering
\includegraphics[width=0.5\textwidth]{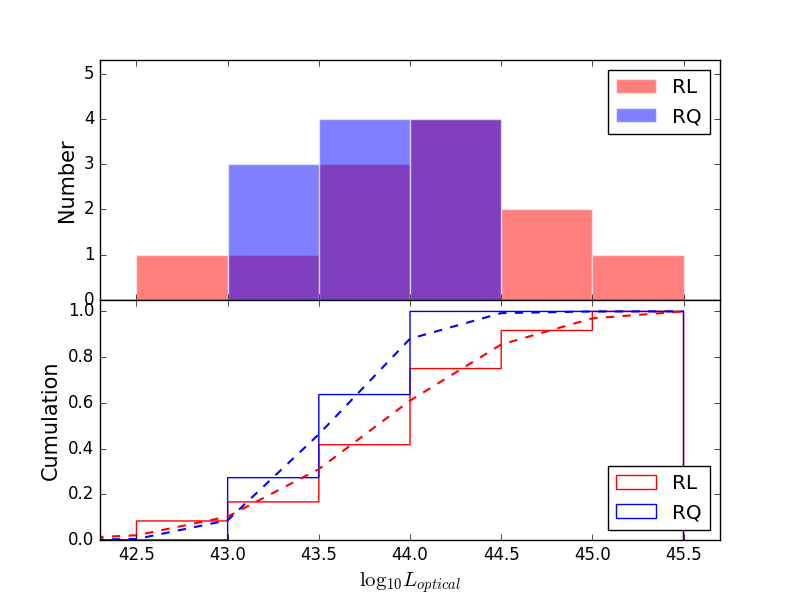}
\caption{\footnotesize Optical luminosity on B-band distribution (top panel) and cumulative distribution (bottom panel) of the RL (red) and RQ (blue) sources as in previous figure using a 0.5 bin width.}
\label{radio+LB}
\end{figure}

The flux densities and luminosities on radio and optical bands, radio loudness, and radio type of the radio sample are reported in Table ~\ref{radio}. The mean values of redshift, black hole mass, radio luminosity at 5 GHz, and optical luminosity on B-band of the RL and RQ subsamples are presented in Table ~\ref{subsample}. Their number and cumulative distributions are shown in Figs. ~\ref{radio+z}, ~\ref{radio+BHmass}, ~\ref{radio+LR}, and ~\ref{radio+LB} respectively.

\setcounter{table}{5}

\begin{table}
\caption{Mean values of redshift, black hole mass, radio luminosity at 5 GHz, and optical luminosity on B-band of the RL and RQ subsample}
\label{subsample}
\centering
\resizebox{0.5\textwidth}{!}{
\begin{tabular}{cccccc}
\hline
\hline
Subsample & Number & Redshift & $\log_{10} (M_{BH} / M_{\odot})$ & $\log_{10} L_{radio}$ & $\log_{10} L_{optical}$ \\
\hline
All & 23 & 0.14 & 6.71 & 39.55 & 43.94 \\
RL & 12 & 0.20 & 6.82 & 40.15 & 44.07 \\
RQ & 11 & 0.06 & 6.59 & 38.90 & 43.79 \\
\hline
\end{tabular}
}
\end{table}

We used the two-sample Kolmogorov-Smirnov (K-S) test to examine whether the parent population of the RL and RQ subsample is the same. The null hypothesis is that two distributions originated from the same population of sources. We applied the rejection of the null hypothesis at a 95$\%$ confidence level corresponding to a value of $p \leq 0.05$. The K-S tests of redshift ($p = 1.8 \times 10^{-3}$) and radio luminosity ($p = 2.0 \times 10^{-4}$) suggest that the RL and RQ NLS1s in the sample have different populations. However, the K-S tests of black hole mass ($p = 0.47$) and optical luminosity ($p = 0.055$) argue that these RL and RQ sources have the same origin.

This result is expected. The different redshift distribution may be caused by a selection effect. RQ sources are faint, so they can be observed only at low redshift due to the flux density limit of radio surveys. Once the redshift increases, only RL sources with high luminosity can be detected. The different radio luminosity distribution may be related to the presence of radio jets. RL NLS1s generally harbor relativistic jets, instead jets in RQ NLS1s are usually weak or absent. However, the optical luminosity and black hole mass distributions have the same origin, the possible reason is that the optical luminosity is emitted by the accretion disk and the black hole mass is derived from the optical continuum in this analysis. This result also implies that the presence of relativistic jets is not related to the central black hole mass.

The cumulative distributions of the radio sample show as well that RL sources tend to have higher redshift, a more massive black hole, and higher radio and optical luminosity than RQ sources on average. This may be related to the jet activity. RL NLS1s usually harbor relativistic jets emitting low-energy synchrotron radiation (radio emission) and high-energy inverse-Compton radiation (up to $\gamma$-ray energy). Instead jets are probably weak or absent in RQ NLS1s. The mass accretion rate of sources with jets is higher than that of sources without jets. Since the presence of jets enhances angular momentum transport, matter in the accretion disk loses angular momentum and falls into the central black hole faster. This can greatly increases the black hole mass growth rate, therefore RL sources have larger black hole masses and higher luminosities compared to RQ sources \citep{2008MNRAS.386..989J}. Finally, we remark that these results should be taken with cautions because of the small radio sample used in these analyses.

\section{Summary}

In this work, we exploited the optical spectra from the 6dFGS DR3, which is currently the most extensive spectroscopic survey available in the southern hemisphere, to perform the first systematic selection of NLS1s in this sky region. The NLS1 sample was selected from this survey and flux calibration for the optical spectra in the sample was derived. The luminosity correlations of L(H$\beta$) - $\lambda$L$_{\lambda}$(5100$\AA$) and L([O III]) - $\lambda$L$_{\lambda}$(5100$\AA$) were found. The central black hole mass and the Eddington ratio were estimated for each target. In addition, the radio counterparts were found for some sources by the radio surveys covering the southern hemisphere. The main results of this paper are summarized as follows.

1. According to the criteria of 600 km s$^{-1}$ $<$ FWHM(H$\beta$) $<$ 2200 km s$^{-1}$ and a flux ratio of [O III] / H$\beta$ $<$ 3, as well as considering the visibility of Fe II multiplets and H$\beta$ line profiles in the optical spectra, we created a new accurate sample of 167 NLS1s from the 6dFGS in the southern hemisphere.

2. The flux-calibrated spectra for these 167 NLS1s in the sample, which were not provided by the 6dFGS, were obtained. To evaluate the reliability of the flux calibration, the relation between the mean flux at 5100 $\AA$ and the magnitude on B-band was calculated for the 6dFGS sample, and the result is in good agreement with the SDSS NLS1 sample from \citet{Cracco2016}. We further compared 15 flux-calibrated spectra in the 6dFGS sample with the new observations specifically carried out at the Asiago Observatory, and confirmed that our flux calibration of the 6dFGS spectra on the H$\beta$ line, [O III] line, and 5100 $\AA$ continuum regions is consistent.

3. The monochromatic luminosity at 5100 $\AA$, and the luminosities of H$\beta$ and [O III] lines were estimated. Strong correlations of L(H$\beta$) - $\lambda$L$_{\lambda}$(5100$\AA$) and L([O III]) - $\lambda$L$_{\lambda}$(5100$\AA$) could be confirmed.

4. The mass of central black hole and the Eddington ratio were calculated for each target, and found to lie in the ranges of $M_{BH} \sim 10^{5.91-7.89} M_{\odot}$ and $L_{bol} / L_{Edd} \sim 0.07-5.35$ respectively, with average values of $M_{BH} = 10^{6.93} M_{\odot}$ and $L_{bol} / L_{Edd} = 0.96$, which are typical values for NLS1s. This result confirms that NLS1s have lower black hole mass and higher Eddington ratio than BLS1s.

5. Of the 167 NLS1s in the 6dFGS sample, 23 (13.8$\%$) sources were found to have associated radio counterparts, including 12 (7.0$\%$) RL NLS1s and 11 RQ NLS1s. RL sources tend to have higher redshift, a more massive black hole, and higher radio and optical luminosities than RQ sources.

Our conclusions increase the number of NLS1s and confirm some well known properties of this peculiar class of AGN. However, the number of NLS1s in the 6dFGS sample, in particular radio sources, is still limited. Further research with larger samples, and higher resolution and sensitivity observations will be necessary to understand the physical mechanism and evolution of NLS1s, with respect to other types of AGN. We expect that the investigation of these sources with high performance observation facilities, located in the southern sites, will help to clarify many of the fundamental questions that have not yet been solved on this class of galaxies.

\begin{acknowledgements}

We thank the anonymous referee for suggestions leading to the improvement of this work.

JHFan's work is partially supported by the National Natural Science Foundation of China (NSFC 11733001, U1531245).

We thank the Final Data Release (DR3) of the Six-degree Field Galaxy Survey (6dFGS).

Funding for the Sloan Digital Sky Survey IV has been provided by the Alfred P. Sloan Foundation, the U.S. Department of Energy Office of Science, and the Participating Institutions. SDSS-IV acknowledges support and resources from the Center for High-Performance Computing at the University of Utah. The SDSS web site is www.sdss.org. SDSS-IV is managed by the Astrophysical Research Consortium for the Participating Institutions of the SDSS Collaboration including the Brazilian Participation Group, the Carnegie Institution for Science, Carnegie Mellon University, the Chilean Participation Group, the French Participation Group, Harvard-Smithsonian Center for Astrophysics, Instituto de Astrof\'isica de Canarias, the Johns Hopkins University, Kavli Institute for the Physics and Mathematics of the Universe (IPMU) / University of Tokyo, Lawrence Berkeley National Laboratory, Leibniz Institut f\"ur Astrophysik Potsdam (AIP), Max-Planck-Institut f\"ur Astronomie (MPIA Heidelberg), Max-Planck-Institut f\"ur Astrophysik (MPA Garching), Max-Planck-Institut f\"ur Extraterrestrische Physik (MPE), National Astronomical Observatories of China, New Mexico State University, New York University, University of Notre Dame, Observat\'ario Nacional / MCTI, the Ohio State University, Pennsylvania State University, Shanghai Astronomical Observatory, United Kingdom Participation Group, Universidad Nacional Aut\'onoma de M\'exico, University of Arizona, University of Colorado Boulder, University of Oxford, University of Portsmouth, University of Utah, University of Virginia, University of Washington, University of Wisconsin, Vanderbilt University, and Yale University.

This research has made use of the NASA/IPAC Extragalactic Database (NED), which is operated by the Jet Propulsion Laboratory, California Institute of Technology, under contract with the National Aeronautics and Space Administration.

This paper is based on observations collected with the 1.22m Galileo telescope and 1.82m Copernico telescopes of the Asiago Astrophysical Observatory, operated by the Department of Physics and Astronomy "G. Galilei" of the University of Padova and INAF - Osservatorio Astronomico di Padova.

This research has made use of the VizieR catalog access tool, CDS, Strasbourg, France. The original description of the VizieR service was published in A\&AS 143, 23.

\end{acknowledgements}

\bibliographystyle{aa}
\bibliography{bibliography}

\onecolumn

\setcounter{table}{0}

\small
\centering
\begin{longtable}{ccccccc}
\caption{\label{All1} Coordinates, redshift, luminosity distance, magnitude on B-band, and mean flux density at 5100 $\AA$ in the 6dFGS NLS1 sample.} \\
\hline
\hline
Name & RA & Dec & Redshift & $D_{L}$ & B & Flux(5100$\AA$) \\
- & (hh:mm:ss) & (dd:mm:ss) & - & (Mpc) & (mag) & (10$^{-17}$ erg s$^{-1}$ cm$^{-2}$ $\AA^{-1}$) \\
\hline
\endfirsthead
\caption{continued.} \\
\hline
\hline
Name & RA & Dec & Redshift & $D_{L}$ & B & Flux(5100$\AA$) \\
- & (hh:mm:ss) & (dd:mm:ss) & - & (Mpc) & (mag) & (10$^{-17}$ erg s$^{-1}$ cm$^{-2}$ $\AA^{-1}$) \\
\hline
\endhead
\hline
\endfoot
6dFGS gJ000040.3-054101 & 00:00:40.26 & -05:41:0.9 & 0.0939 & 430.38 & 17.765 & 41.35 $\pm$ 16.79 \\
6dFGS gJ001302.3-323830 & 00:13:2.25 & -32:38:29.9 & 0.2593 & 1313.9 & 18.577 & 30.25 $\pm$ 2.75 \\
6dFGS gJ001521.6-150951 & 00:15:21.63 & -15:09:51.0 & 0.0788 & 357.38 & 18.717 & 15.33 $\pm$ 3.98 \\
6dFGS gJ001641.9-312657 & 00:16:41.86 & -31:26:56.9 & 0.3604 & 1921.63 & 18.063 & 33.69 $\pm$ 3.69 \\
6dFGS gJ002121.6-205018 & 00:21:21.55 & -20:50:17.6 & 0.183 & 887.91 & 17.697 & 35.59 $\pm$ 3.2 \\
6dFGS gJ002249.2-103956 & 00:22:49.16 & -10:39:55.8 & 0.4142 & 2263.79 & 18.425 & 23.21 $\pm$ 3.42 \\
6dFGS gJ003000.5-202856 & 00:30:0.52 & -20:28:56.2 & 0.2889 & 1486.83 & 18.652 & 5.8 $\pm$ 1.57 \\
6dFGS gJ003211.0-324205 & 00:32:10.95 & -32:42:5.1 & 0.3539 & 1881.14 & 18.667 & 18.87 $\pm$ 2.22 \\
\hline
\end{longtable}
\tablefoot{Column 1) Name in the 6dFGS. 2) Right Ascension. 3) Declination. 4) Redshift. 5) Luminosity distance. 6) Magnitude on B-band. 7) Mean flux density at 5100 $\AA$.}

\small
\centering
\begin{longtable}{cccccc}
\caption{\label{All2} FWHM and flux of H$\beta$ and [O III] lines and their flux ratio in the 6dFGS NLS1 sample.} \\
\hline
\hline
Name & FWHM(H$\beta$) & FWHM([O III]) & Flux(H$\beta$) & Flux([O III]) & Flux ratio [O III]/H$\beta$ \\
- & (km s$^{-1}$) & (km s$^{-1}$) & (10$^{-17}$ erg s$^{-1}$ cm$^{-2}$) & (10$^{-17}$ erg s$^{-1}$ cm$^{-2}$) & - \\
\hline
\endfirsthead
\caption{continued.} \\
\hline
\hline
Name & FWHM(H$\beta$) & FWHM([O III]) & Flux(H$\beta$) & Flux([O III]) & Flux ratio [O III]/H$\beta$ \\
- & (km s$^{-1}$) & (km s$^{-1}$) & (10$^{-17}$ erg s$^{-1}$ cm$^{-2}$) & (10$^{-17}$ erg s$^{-1}$ cm$^{-2}$) & - \\
\hline
\endhead
\hline
\endfoot
6dFGS gJ000040.3-054101 & 1012.5 $\pm$ 93.0 & 714.9 $\pm$ 33.0 & 1428.91 $\pm$ 80.23 & 1143.99 $\pm$ 53.88 & 0.8 $\pm$ 0.06 \\
6dFGS gJ001302.3-323830 & 2189.9 $\pm$ 124.4 & 586.5 $\pm$ 52.9 & 1671.05 $\pm$ 29.84 & 503.24 $\pm$ 18.57 & 0.3 $\pm$ 0.01 \\
6dFGS gJ001521.6-150951 & 1120.1 $\pm$ 85.4 & 362.5 $\pm$ 36.3 & 1055.21 $\pm$ 40.33 & 435.65 $\pm$ 23.56 & 0.41 $\pm$ 0.03 \\
6dFGS gJ001641.9-312657 & 2237.4 $\pm$ 72.5 & 950.4 $\pm$ 212.2 & 2106.58 $\pm$ 31.16 & 123.75 $\pm$ 17.77 & 0.06 $\pm$ 0.01 \\
6dFGS gJ002121.6-205018 & 1533.2 $\pm$ 62.2 & 909.7 $\pm$ 137.3 & 2192.49 $\pm$ 35.24 & 212.2 $\pm$ 22.52 & 0.1 $\pm$ 0.01 \\
6dFGS gJ002249.2-103956 & 1852.7 $\pm$ 71.3 & 761.1 $\pm$ 42.0 & 1763.16 $\pm$ 28.29 & 912.99 $\pm$ 20.65 & 0.52 $\pm$ 0.01 \\
6dFGS gJ003000.5-202856 & 1172.9 $\pm$ 58.7 & 379.6 $\pm$ 50.6 & 342.54 $\pm$ 13.22 & 159.86 $\pm$ 10.4 & 0.47 $\pm$ 0.04 \\
6dFGS gJ003211.0-324205 & 1124.7 $\pm$ 73.5 & 477.7 $\pm$ 17.4 & 886.78 $\pm$ 20.02 & 376.26 $\pm$ 15.64 & 0.42 $\pm$ 0.02 \\
\hline
\end{longtable}
\tablefoot{Column 1) Name in the 6dFGS. 2) FWHM of H$\beta$ line fitted with three Gaussians. 3) FWHM of [O III] line fitted with two Gaussians. 4) Flux of H$\beta$ line. 5) Flux of [O III] line. 6) Flux ratio of [O III] / H$\beta$.}

\setcounter{table}{3}

\small
\centering
\begin{longtable}{cccccc}
\caption{\label{All3} Luminosity at 5100 $\AA$, luminosities of H$\beta$ and [O III] lines, mass of central black hole, and Eddington ratio in the 6dFGS NLS1 sample.} \\
\hline
\hline
Name & L(5100$\AA$) & L(H$\beta$) & L([O III]) & $M_{BH}$ & $\epsilon_{Edd}$ \\
- & (10$^{43}$ erg s$^{-1}$) & (10$^{40}$ erg s$^{-1}$) & (10$^{40}$ erg s$^{-1}$) & (10$^{6}$ $M_{\odot}$) & - \\
\hline
\endfirsthead
\caption{continued.} \\
\hline
\hline
Name & L(5100$\AA$) & L(H$\beta$) & L([O III]) & $M_{BH}$ & $\epsilon_{Edd}$ \\
- & (10$^{43}$ erg s$^{-1}$) & (10$^{40}$ erg s$^{-1}$) & (10$^{40}$ erg s$^{-1}$) & (10$^{6}$ $M_{\odot}$) & - \\
\hline
\endhead
\hline
\endfoot
6dFGS gJ000040.3-054101 & 2.34 $\pm$ 0.86 & 31.67 $\pm$ 1.78 & 25.35 $\pm$ 1.19 & 2.3 $\pm$ 0.73 & 0.71 $\pm$ 0.34 \\
6dFGS gJ001302.3-323830 & 21.81 $\pm$ 2.77 & 345.16 $\pm$ 6.16 & 103.95 $\pm$ 3.84 & 15.08 $\pm$ 3.86 & 1.0 $\pm$ 0.29 \\
6dFGS gJ001521.6-150951 & 1.14 $\pm$ 0.31 & 16.13 $\pm$ 0.62 & 6.66 $\pm$ 0.36 & 3.37 $\pm$ 0.99 & 0.23 $\pm$ 0.09 \\
6dFGS gJ001641.9-312657 & 41.47 $\pm$ 8.2 & 930.74 $\pm$ 13.77 & 54.68 $\pm$ 7.85 & 18.21 $\pm$ 4.95 & 1.58 $\pm$ 0.53 \\
6dFGS gJ002121.6-205018 & 13.75 $\pm$ 1.53 & 206.82 $\pm$ 3.32 & 20.02 $\pm$ 2.12 & 8.85 $\pm$ 2.24 & 1.08 $\pm$ 0.3 \\
6dFGS gJ002249.2-103956 & 43.18 $\pm$ 10.6 & 1081.12 $\pm$ 17.35 & 559.82 $\pm$ 12.66 & 17.08 $\pm$ 4.83 & 1.75 $\pm$ 0.66 \\
6dFGS gJ003000.5-202856 & 3.65 $\pm$ 2.02 & 90.6 $\pm$ 3.5 & 42.28 $\pm$ 2.75 & 10.22 $\pm$ 3.94 & 0.25 $\pm$ 0.17 \\
6dFGS gJ003211.0-324205 & 17.37 $\pm$ 4.53 & 375.46 $\pm$ 8.48 & 159.31 $\pm$ 6.62 & 12.52 $\pm$ 3.54 & 0.96 $\pm$ 0.37 \\
\hline
\end{longtable}
\tablefoot{Column 1) Name in the 6dFGS. 2) Luminosity at 5100 $\AA$. 3) Luminosity of H$\beta$ line. 4) Luminosity of [O III] line. 5) Mass of central black hole. 6) Eddington ratio.}

\tiny
\centering
\begin{landscape}
\begin{longtable}{ccccccccccccc}
\caption{\label{radio} Flux densities and luminosities on radio and optical bands, radio loudness, and radio type in the radio sample.} \\
\hline
\hline
Name & S843MHz & S1.4GHz & S5GHz & S8.4GHz & S22GHz & F(4400$\AA$) & L(Radio) & L(4400$\AA$) & $R_{L}$ & $R_{L}$ & $R_{L}$ & Type \\
- & (mJy) & (mJy) & (mJy) & (mJy) & (mJy) & (10$^{-17}$ erg s$^{-1}$ cm$^{-2}$ $\AA^{-1}$) & (10$^{38}$ erg s$^{-1}$) & (10$^{43}$ erg s$^{-1}$) & ($\alpha$ = 0) & ($\alpha$ = 0.5) & ($\alpha$ = 1) & - \\
\hline
\endhead
\hline
\endfoot
6dFGS gJ012237.5-264646 & - & 3.8 $\pm$ 0.5 & - & - & - & 4.0 $\pm$ 1.97 & 332.24 $\pm$ 43.72 & 10.99 $\pm$ 5.42 & 147.1 & 77.9 & 41.2 & RL \\
6dFGS gJ042256.6-185442 & - & 2.8 $\pm$ 0.5 & - & - & - & 78.9 $\pm$ 9.66 & 3.89 $\pm$ 0.69 & 3.44 $\pm$ 0.42 & 5.5 & 2.9 & 1.5 & RQ \\
6dFGS gJ043622.3-102234 & - & 17.0 $\pm$ 0.7 & - & - & - & 869.95 $\pm$ 120.41 & 6.81 $\pm$ 0.28 & 10.95 $\pm$ 1.52 & 3.0 & 1.6 & 0.8 & RQ \\
6dFGS gJ045230.1-295335 & - & 9.5 $\pm$ 0.5 & - & - & - & 16.29 $\pm$ 1.94 & 343.15 $\pm$ 18.06 & 18.49 $\pm$ 2.21 & 90.3 & 47.8 & 25.3 & RL \\
6dFGS gJ054914.9-242552 & - & 2.4 $\pm$ 0.5 & - & - & - & 118.9 $\pm$ 12.51 & 1.6 $\pm$ 0.33 & 2.49 $\pm$ 0.26 & 3.1 & 1.7 & 0.9 & RQ \\
6dFGS gJ062233.5-231742 & - & 4.3 $\pm$ 0.5 & - & - & - & 238.19 $\pm$ 26.21 & 2.0 $\pm$ 0.23 & 3.48 $\pm$ 0.38 & 2.8 & 1.5 & 0.8 & RQ \\
6dFGS gJ084628.7-121409 & - & 15.3 $\pm$ 0.7 & - & - & - & 60.24 $\pm$ 6.78 & 63.74 $\pm$ 2.92 & 7.89 $\pm$ 0.89 & 39.3 & 20.8 & 11.0 & RL \\
6dFGS gJ095219.1-013644 & - & 62.2 $\pm$ 1.9 & - & - & - & 634.94 $\pm$ 104.25 & 7.56 $\pm$ 0.23 & 2.43 $\pm$ 0.4 & 15.2 & 8.0 & 4.2 & RQ \\
\hline
\end{longtable}
\tablefoot{Column 1) Name in the 6dFGS. 2) Flux density detected by the SUMSS at 843 MHz. 3) Flux density detected by the NVSS at 1.4 GHz. Columns 4), 5), 6) Flux densities detected by the AT20G at 5 GHz, 8.4 GHz, and 22 GHz respectively. 7) Optical flux density on B-band. 8) Radio luminosity at 5 GHz. 9) Optical luminosity on B-band. Columns 10), 11), 12) Radio-loudness assuming a spectral index of $\alpha$=0, $\alpha$=0.5, and $\alpha$=1 respectively. 13) Radio type according to the R$_{L}$($\alpha$=0.5).}
\end{landscape}

\end{document}